\documentclass[11pt,reqno]{article}
\usepackage{amsmath,hyperref,amssymb, amsthm}
\usepackage{authblk}
\usepackage[all]{xy}
\usepackage{graphicx,url}
\usepackage{color}
\usepackage{slashed}

\newcommand{\be}{\begin{eqnarray}}
\newcommand{\ee}{\end{eqnarray}}
\newcommand{\eeq}{\end{equation}}
\newcommand{\beq}{\begin{equation}}

\allowdisplaybreaks \numberwithin{equation}{section}

\DeclareSymbolFont{AMSa}{U}{msa}{m}{n}
\DeclareSymbolFont{AMSb}{U}{msb}{m}{n}
\DeclareMathSymbol{\fieldR}{\mathalpha}{AMSb}{"52}

\newcommand{\ZZ}{\mathbb{Z}}
\newcommand{\RR}{\mathbb{R}}

\def\beq{\begin{equation}}
\def\eeq{\end{equation}}
\def\bea{\begin{eqnarray}}
\def\eea{\end{eqnarray}}

\def\<{\langle}

\title{Snowmass White Paper: Moonshine}
\author[1,2]{Sarah M. Harrison}
\author[3]{Jeffrey A. Harvey}
\author[4]{Natalie M. Paquette}

{\small 	
\affil[1]{\small Department of Mathematics and Statistics, McGill University, Montreal, QC, Canada}
	\affil[2]{\small Department of Physics, McGill University, Montreal, QC, Canada}
	\affil[3]{\small Enrico Fermi Institute and Department of Physics, University of Chicago \\\newline
5640 Ellis Ave., Chicago IL 60637, USA}
	\affil[4]{Department of Physics, University of Washington, Seattle, WA, 98195, USA}}
\date{}

\begin{document}
\nocite{*}
\maketitle
\abstract{We present a brief overview of Moonshine with an emphasis on connections to physics. Moonshine collectively refers to a set of phenomena connecting group theory, analytic number theory, and vertex operator algebras or conformal field theories. Modern incarnations of Moonshine arise in various BPS observables in string theory and, via dualities, invariants in enumerative geometry. We survey old and new developments, and highlight some of the many open questions that remain.}

\vspace{2cm}
\newpage
\tableofcontents

\section{Introduction}

This paper contains a brief review of various types of moonshine starting with the
earliest example of monstrous moonshine. Other reviews of moonshine with various points of view can be found in \cite{Anagiannis:2018jqf, MR3375653,duncan2019monster, Kachru:2016nty} and the book \cite{MR2257727} describes moonshine as of 2006.

Monstrous moonshine had its origins in observations of McKay and Thompson of
connections between the representation theory of the Monster ($\mathbb{M}$), the largest sporadic finite simple group, and the elliptic modular invariant
\beq
J(\tau)= q^{-1} +196884 q + 21493760 q^2 + \cdots \, .
\eeq
Here $q=e^{2 \pi i \tau}$ and $\tau$ is a complex parameter taking values in the upper half plane ${\rm Im}(\tau) >0$. $J(\tau)$ is the unique function with a simple pole only at $q=0$ with residue one and vanishing constant term that is invariant
under modular transformations $\tau \rightarrow (a \tau +b)/(c \tau +d)$ with $a,b,c,d \in \ZZ$ and $ad-bc=1$. The idea of a connection between modular functions and the representation theory of the Monster group is so outlandish that it was referred to as moonshine by Conway and Norton \cite{CN}. A connection between these two areas of mathematics eventually emerged with a striking connection to physics: consistent two-dimensional conformal field theories (CFTs) have partition functions which can be computed as path integrals on a two-dimensional torus and are modular invariant functions as a consequence of invariance under global diffeomorphisms acting on the torus. Thus
the construction of a (holomorphic) CFT with Monster symmetry in \cite{MR747596,MR996026}
(or more precisely a vertex operator algebra (VOA)) led to an explanation of this bizarre connection. A physical perspective on this construction was given in \cite{Dixon:1988qd}.  We now understand that it is easy to find more connections of this sort. One can start with a positive definite lattice $L$ with a large automorphism group $G$ and construct a lattice VOA whose partition function will have modular properties and have coefficients that can be written as dimensions of representations of a group closely related to $G$. However, the word moonshine, in the words of R. Borcherds, ``...should only be applied to things that are weird and special; if there are an infinite number of examples of something, then it is not moonshine." The thing that makes Monstrous moonshine special is a connection to genus zero subgroups of $SL(2,\RR)$ as explained later in this review.

The classic era of moonshine involved the study of Monstrous moonshine and a supersymmetric variant known as Conway moonshine \cite{MR781381,MR2352133,Duncan:2014eha}. A new era of moonshine began in 2010 with the observation in \cite{Eguchi:2010ej} of a surprising connection between the elliptic genus of $K3$ surfaces and the representation theory of another sporadic group, the Mathieu group $M_{24}$. This relation involves a new kind of modular object, a mock
modular form. This relation was soon extended to umbral moonshine \cite{UM, UMNL,Cheng:2017usy}, which conjectured relationships between a set of $23$ distinguished mock modular forms and $23$ finite groups which arise from the automorphism groups of Niemeier lattices. The main conjecture of Mathieu and umbral moonshine was proven in \cite{MR3539377,MR3433373} but as we explain below, there is still a great deal of mystery surrounding the origins of Mathieu and umbral moonshine.  More recently there have been additional new types of moonshine: for the O'Nan sporadic group \cite{MR4291251} and penumbral moonshine \cite{penumbral} which encompasses moonshine for the Thompson group \cite{Harvey:2015mca} as a special case, much as umbral moonshine generalizes Mathieu moonshine. Although the connections of these new types of moonshine to physics are at the moment unknown, they do deserve to be called moonshine in that they are all special and finite in number and this again arises through connections to genus zero subgroups of $SL(2, \RR)$ \cite{UMNL,Cheng:2016klu, penumbral}.

\section{Connections to CFT/VOA}

Monstrous moonshine had its origins in observations of J. McKay and J. Thompson of
connections between the representation theory of the Monster, the largest sporadic finite simple group, and the elliptic modular invariant $J(\tau)$. 
For example, the coefficient $196884=1+196883$ is the sum of the dimensions of the two smallest irreducible representations of $\mathbb{M}$. These observations suggested the existence of an infinite-dimensional graded vector space, $V^\natural$, carrying representations of $\mathbb{M}$, and a generalization $T_g(\tau)$ of $J(\tau)$, known as McKay-Thompson series. These series are constructed by replacing the dimensions of representations by characters of group elements of $\mathbb{M}$. Conway and Norton organized and generalized these observations and made several conjectures, the most important of which \cite{MR554402} was that each of the $T_g(\tau)$ (which depend only on the conjugacy class of $g$) \footnote{``Generalized moonshine'' for a larger class of functions $T_{g, h}(\tau)$ labeled by commuting pairs $g, h \in \mathbb{M}$ was conjectured by Norton \cite{norton1987moonshine, norton2001moonshine} and proved by Carnahan \cite{carnahan2010generalized, Carnahan:2009mjm, Carnahan:2012gx}, who constructed new BKMs to establish their genus zero properties. The natural physical origin of these functions is as twisted sector partition functions of orbifolds of $V^{\natural}$; see e.g. \cite{Dixon:1988qd,tuite1992monstrous, tuite2010monstrous} for details.} were invariant under a special class of subgroups of $SL(2,\mathbb{R})$ known as genus zero subgroups because the quotient of the upper half plane by these groups can be given the structure of a genus zero Riemann surface.  The proof of this genus zero conjecture involved the construction of a Vertex Operator Algebra (VOA) on $V^\natural$ by Frenkel, Lepowsky and Meurman (FLM) \cite{MR747596,MR996026} following on the definition of a Vertex Algebra by Borcherds \cite{MR843307} and followed by a tour de force proof by Borcherds of the genus zero property \cite{MR1172696}. Borcherds defined and employed a new class of algebras known as Borcherds-Kac-Moody (BKM) algebras, see \cite{MR1660651} for a discussion of his work.

Remarkably these ideas and techniques have a close relationship to physics. In particular, a VOA is, roughly speaking, equivalent to the chiral part of a two-dimensional conformal field theory (CFT) and the construction of FLM can be viewed as the first example of an orbifold CFT, an idea which was developed independently in the context of string compactifications and then eventually abstracted into a general mechanism for constructing new CFTs from old by gauging global symmetries. Recently, a physical explanation for the genus zero property of Monstrous moonshine has been proposed in the context of a family of special heterotic string compactifications including $V^{\natural}$ or its orbifolds on the worldsheet \cite{Paquette:2016xoo, Paquette:2017xui}; there, the genus zero property is shown to follow from self-duality of the models under T-dualities.

Moonshine for another sporadic group, the Conway group $Co_1$ arises in a superconformal field theory with central charge $c=12$ \cite{MR781381,MR2352133,Duncan:2014eha}. 
The corresponding McKay-Thompson series are also associated to genus zero subgroups of $SL(2, \mathbb{R})$ \cite{Duncan:2014eha}. 

Moonshine relations between modular functions and other sporadic groups that are involved in the Monster group have also been constructed using ideas from conformal field theory starting with work of H{\"o}hn on the Baby Monster group \cite{hohn}. An important ingredient in his and other constructions is the fact that certain minimal model CFTs and parafermion CFTs can be embedded in the Monster CFT and the discrete symmetries of the minimal model or parafermion CFT then act as elements of the Monster group. For example, the $\mathbb{Z}_2$ symmetry of the $c=1/2$ Ising model acts as an element of the 2A conjugacy class of the Monster, an obervation due to Miyamoto.
Many generalizations of this idea have been worked out, see
\cite{MR1618135,Bae:2020pvv} for further examples and constructions. A similar phenomenon for the Conway CFT is explored in \cite{Cheng:2014owa}.

One of the goals of current research into moonshine is to find CFT/VOA explanations for umbral and penumbral moonshine. The Mathieu moonshine example of umbral moonshine arose through a study of the decomposition of the elliptic genus of $K3$ surfaces into representations of the $N=4$ superconformal algebra that exists for any $K3$ sigma-model.
The multiplicities of massive $N=4$ characters are captured by a weight $1/2$ mock modular form with $q$ expansion
\beq
H^{(2)}(\tau)= 2( -q^{-1/8} + 45 q^{7/8} + 231 q^{15/8} + \cdots).
\eeq
Additionally, associated to each $g\in M_{24}$ is a McKay-Thompson series $H^{(2)}_g(\tau)$, also a weight $1/2$ mock modular form for a subgroup of $SL(2,\mathbb Z)$, whose form was originally conjectured in \cite{Cheng:2010pq,Gaberdiel:2010ch,Gaberdiel:2010ca,Eguchi:2010fg}.

The occurrence of dimensions of $M_{24}$ irreducible representations would not be surprising if there existed $K3$ sigma-models with $M_{24}$ group actions that preserve the $N=4$ algebra. However both classical algebro-geometric analysis \cite{MR958597} and quantum analysis of $K3$ sigma models \cite{Gaberdiel:2011fg} shows that no such $K3$ sigma-model exists\footnote{One may also study global symmetries that preserve only the $\mathcal{N}=(4,1)$ superconformal symmetry required for the definition of the elliptic genus and its decomposition into $\mathcal{N}=4$ characters, rather than the full $\mathcal{N}=(4,4)$ algebra \cite{Harvey:2020jvu}; although the resulting symmetry groups are larger, this has not led to $M_{24}$ symmetry.}. The result of the classification of \cite{Gaberdiel:2011fg} is that all symmetry groups of K3 sigma-models are (proper) subgroups of the Conway group, the automorphism group of the Leech lattice. In particular, this implies there exist K3 sigma-models with symmetries which lie outside of $M_{24}$, as well as elements of $M_{24}$ which can never appear as symmetries of K3 sigma models. This result was generalized in \cite{Cheng:2016org} to include singular points in the moduli space of K3 CFTs (which correspond to points of enhanced gauge symmetry in spacetimes), and where the possible twining genera of K3 sigma models were classified and shown to be related to the umbral and Conway moonshine constructions of \cite{Cheng:2014zpa} and \cite{Duncan:2015xoa}, respectively.

Given that the elliptic genus only captures contributions of BPS states, which are independent of the moduli of the theory, one attempt to overcome the problem that no K3 CFT has $M_{24}$ symmetry involves the idea of ``symmetry-surfing'' which involves combining symmetries that appear at different points in the moduli space of $K3$ sigma-models, first explored for Kummer surfaces in  \cite{Taormina:2011rr,Taormina:2013jza}, and moving away from the Kummer locus in \cite{Gaberdiel:2016iyz}. A similar idea was investigated for UV Landau-Ginzburg models which flow to K3 CFTs in the IR in \cite{Cheng:2015rby}. The elliptic genus of $K3$ can also be viewed as the partition function of a ``half-twisted" topological field theory and is closely related \cite{Kapustin:2005pt} to another mathematical construction known as the chiral de Rham complex. It has been suggested that this might be the correct arena to understand Mathieu moonshine, see for example \cite{Wendland:2017eiw}.

The modular objects appearing in umbral moonshine are mock modular forms of a type discovered by Ramanujan \cite{rama} and put into a modern mathematical framework by Zwegers \cite{zwegers}. Mock modularity appears in computations of the elliptic genus of non-compact sigma models \cite{Eguchi:2010cb,Troost:2010ud} so it is natural to look for a CFT involving both a $K3$ surface and a non-compact sigma model in order to explain the mock modularity of the functions appearing in umbral moonshine. Such models appear in the description of fivebranes wrapping $K3$ surfaces \cite{Harvey:2013mda} and in the description of string theory near an ADE singularity \cite{Cheng:2014zpa}. Analysis of these CFTs lead to intriguing connections to the mock modular forms of umbral moonshine but have not yet led to an explicit construction of the umbral moonshine modules. 
Another attempt to understand Mathieu moonshine involves exploiting connections between
certain $K3$ sigma-models and the Conway moonshine CFT \cite{Duncan:2015xoa,Taormina:2017zlm}.\footnote{An analogous story for $T^4$ sigma models was explored in \cite{Volpato:2014zla,Anagiannis:2020hkk}.}

However, the existence of generalized Mathieu \cite{Gaberdiel:2012gf} and umbral moonshines \cite{Cheng:2016nto} suggests that there should eventually be a CFT/VOA-related explanation for these phenomena. The case of umbral moonshine corresponding to Niemeier root system $3E_8$ was given a VOA interpretation in \cite{Duncan:2014tya}. Umbral moonshine can be formulated in terms of a specific set of optimal mock Jacobi forms which descend from meromorphic Jacobi forms by a procedure discussed in \cite{Dabholkar:2012nd}. There exist VOA  constructions (based on free fields) of the (twined) meromorphic Jacobi forms of umbral moonshine corresponding to Niemeier root systems $4A_6$ and $2A_{12}$ \cite{Duncan:2017bhh} and $4D_6, 3D_8, 2D_{12},$ and $D_{24}$ \cite{Cheng:2017grj}, thus furnishing a solution to the module problem for these cases. Additionally, in  \cite{Anagiannis:2017src}, the relationship between umbral moonshine, the K3 elliptic genus, and the Conway moonshine module was exploited to construct a module for the case of umbral moonshine corresponding to the Niemeier lattice with root system $6D_4$. However, many cases, including the original case of Mathieu moonshine corresponding to root system $24A_1$ are still missing module constructions. Furthermore, a uniform construction of umbral moonshine modules is still lacking, as none of the existing constructions can be generalized to all cases at once.

\section{Connections to string theory}

Moonshine's close historical connections to VOA (and consequently, CFT) imply connections to distinguished worldsheet string theories. Less immediate are the many beautiful intersections of moonshine with spacetime string theory, which may provide an arena for resolving the outstanding mysteries of moonshine, or provide a setting for novel applications of number theory and large sporadic groups to physics.

One theme of these explorations is the fact that BPS state counting functions are automorphic objects. For judiciously chosen compactifications, these objects may coincide with, or otherwise be related to, moonshine functions. The central role of mock modular forms in Mathieu and umbral moonshines is suggestive of noncompact systems, such as the near-horizon limit of NS5-branes and little string theories; perturbative BPS states in such systems are counted by the spacetime helicity supertrace, whose decomposition into mock modular forms and relations to moonshine were explored in \cite{Harvey:2013mda, Harvey:2014cva}. By contrast, BPS counting functions for certain compact fivebranes wrapping Calabi-Yau divisors produce skew-holomorphic Jacobi forms with moonshine properties \cite{Denef:2007vg,Cheng:2017dlj}. The new supersymmetric index for the heterotic string on $K3 \times T^2$, and its type II duals, inherit moonshine features from the $K3$ elliptic genus \cite{Cheng:2013kpa}, which persist in 4d $\mathcal{N}=1$ compactifications with flux \cite{Wrase:2014fja, Paquette:2014rma} and in orbifolds \cite{Datta:2015hza, Chattopadhyaya:2016xpa, Chattopadhyaya:2017zul, Banlaki:2018pcc}, although it is not yet understood whether or how $M_{24}$ acts in these systems. Siegel forms counting 1/4-BPS dyons (black holes, at large charge) in 4d $\mathcal{N}=4$ compactifications from type II on $K3 \times T^2$ and its CHL orbifolds may be decomposed into mock modular forms \cite{Dabholkar:2012nd} and exhibit a variety of (umbral and Conway) moonshine connections \cite{Cheng:2010pq, Paquette:2017gmb, Benjamin:2017rnd, Chattopadhyaya:2017ews, Persson:2013xpa} and novel string dualities \cite{Persson:2015jka}. Furthermore, there are many suggestive connections between dyon counting functions and BKM algebras \cite{Dabholkar:2006xa, Cheng:2008fc, Cheng:2008kt, Govindarajan:2008vi, Govindarajan:2011mp, Govindarajan:2010fu, Govindarajan:2009qt}. (Recently, other number theoretic quantities have been recognized in black hole counting functions which also await explanation \cite{Gunaydin:2019xxl, Kachru:2017eju}).

On another front, special low-dimensional spacetime string theory constructions \cite{Harvey:1987da} have been proposed to explain the genus zero properties of Monstrous \cite{Paquette:2016xoo, Paquette:2017xui} and Conway moonshines \cite{Harrison:2021gnp}, with the associated BKM algebras \cite{Harrison:2020wxl, Harrison:2018joy} playing the role of BPS spectrum-generating algebras. BKM algebras have long been suggested to furnish an algebra of BPS states in string compactifications \cite{Harvey:1995fq, Harvey:1996gc}, and it is a fascinating open possibility that moonshine groups may be best understood as symmetries of BPS algebras. Other low-dimensional string compactifications \footnote{See also the closely related works \cite{Harrison:2016pmb, Kachru:2017ulh}.} naturally enjoy the umbral \cite{Kachru:2016ttg} and Conway \cite{Kachru:2016ttg, Harvey:2017xdt, Baykara:2021ger} groups as subgroups of the duality group, suggesting a governing role for large sporadic groups in organizing spacetime symmetries. The role of mock modular forms in these compactifications is yet to be understood, though a proposal from a D1-NS5 system in 2d appears in \cite{Zimet:2018dev}. 

Finally, in the context of holography, moonshine VOAs have been explored as instances of the AdS$_3$/CFT$_2$ correspondence \cite{Witten:2007kt,Gaiotto:2008jt,Duncan:2009sq}, particularly as putative duals to pure 3d gravity (so-called extremal CFTs) with $l_{AdS} \sim \alpha'^{1/2}$. Many questions about such constructions remain. However, the Rademacher summability properties of moonshine functions (see below) are inspired by, and may presage further connections to, holography \cite{Cheng:2011ay}.

\section{Mathematical connections}
One way to construct a modular form is by a Poincar\'e series, roughly constructed via an infinite sum over $SL(2,\mathbb Z)$ images of a seed function \cite{PoincarFonctionsME}. This sum may have convergence issues depending on the weight of the modular form and form of the seed function, which were resolved in weight zero by a regularization procedure due to Rademacher \cite{Rad}. We refer to these regularized Poincar\'e series as Rademacher sums. However, after regularization, this Rademacher sum may no longer be a modular function, but have mock modular properties (see, e.g., \cite{Cheng:2012qc}).
The obstruction to modularity occurs in what is known as Eichler cohomology and work of Knopp shows that it vanishes when the invariance group imposed on the Rademacher sums has genus zero. See \cite{MR1050700} for an overview of these mathematical developments. 

The connection between Rademacher sums and moonshine phenomena began with \cite{Duncan:2009sq}.  Rademacher sums for other weights have been developed in the mathematical literature, see \cite{Cheng:2012qc} for a review. These have played an important role in the characterization of the mock modular forms of Mathieu and umbral moonshine \cite{Cheng:2011ay,Cheng:2012qc}, as well as the trace functions appearing in penumbral moonshine \cite{penumbral}.

Rademacher sums have also been proposed to have a physical interpretation in terms of the AdS$_3$/CFT$_2$ correspondence \cite{Dijkgraaf:2000fq,Manschot:2007ha}, in which the terms in the sum correspond to saddle points of the gravitational path integral with toroidal boundary condition. This has been developed further in \cite{deBoer:2006vg,Manschot:2007ha,Duncan:2009sq,Murthy:2009dq}. Other applications of Rademacher sums include exact asymptotic expansions for Fourier coefficients of certain modular forms, which have been applied to 2d CFT partitions functions \cite{Alday:2019vdr}, Vafa-Witten invariants \cite{Bringmann:2010sd}, and the entropy of supersymmetric black holes in string theory \cite{Dabholkar:2005by,Dabholkar:2005dt,Dabholkar:2014ema,Ferrari:2017msn,Cardoso:2021gfg}.

While the monstrous moonshine conjectures concerned a relation between the monster group and classical modular functions, more recent incarnations of moonshine---including umbral and penumbral moonshine---involve connections between finite groups and more exotic automorphic forms of varying weights, including mock modular forms, (weak, mock, meromorphic and skew-holomorphic) Jacobi forms, and Siegel modular forms. For a review of the modular objects connected to Mathieu moonshine, see \cite{Cheng:2012rca}.
See \cite{Cheng:2016klu} for the classification of ``optimal" mock Jacobi forms of weight 1, including those which appear in umbral moonshine, and a connection to genus zero groups and \cite{penumbral} for a parallel discussion of skew-holomorphic Jacobi forms and genus zero groups in penumbral moonshine.

There have also been some tantalizing connections between automorphic forms connected with moonshine and invariants natural in enumerative geometry. In the case of K3 surfaces, enumerative invariants have been studied since the work of Yau and Zaslow \cite{Yau:1995mv}, who conjectured that the generating function for  rational curves with $n$ double points on K3 is given by $1/\eta^{24}(\tau)$, equal to the generating function for Euler characteristics of Hilb$^{[n]}$(K3). Since then, further refinements of this formula have been conjectured by considering  Gopakumar-Vafa invariants arising from M theory on $K3 \times T^2$  \cite{Katz:1999xq} as well as motivic stable pairs invariants \cite{Katz:2014uaa}, which reproduce the generating function of Hodge numbers of Hilb$^{[n]}$(K3). In \cite{Katz:2014uaa} the question was posed whether these invariants are naturally related to dimensions of $M_{24}$ representations. Though the answer to this question appears to be negative \cite{Harvey:2017xdt}, in \cite{Cheng:2015kha}, inspired by \cite{Gaberdiel:2011fg}, it was proposed that there is a natural action of subgroups of the Conway group on these invariants. In \cite{Huybrechts:2013iwa}, the classification result of \cite{Gaberdiel:2011fg} was reinterpreted in terms of groups of derived equivalences of K3 surfaces. Finally, the Igusa cusp form $\Phi_{10}$ has been conjectured to reproduce the full Gromov-Witten theory of $K3 \times T^2$ \cite{Oberdieck2016CurveCO}, and been generalized to CHL orbifolds of $K3 \times T^2$ in \cite{Bryan:2018nlv,Fischbach:2020bji}, where a number of the Siegel modular forms arising from second quantized Mathieu moonshine have appeared \cite{Persson:2013xpa}. 

The phenomenon of O'Nan moonshine has been observed to have connections to the arithmetic of elliptic curves \cite{duncan2021nan} as has moonshine for the Thompson group \cite{MR4230542}. We are just beginning to scratch the surface of this intriguing connection between sporadic groups, automorphic forms, and arithmetic geometry.

In our brief description of developments in moonshine we have omitted one aspect of moonshine which at the moment seems far from physics. The representation theory of finite groups can be defined over fields of positive characteristic $p$. The study of these representations is known as modular representation theory. In the papers \cite{MR1372729,MR1390654} Ryba and Borcherds observed some very interesting behavior in the $p$-modular behavior of the module $V^\natural$ of the Monster VOA. One of their main conjectures was recently proved by Carnahan \cite{MR3941099}. A nice overview of these developments as well as observations and speculations regarding connections between modular moonshine and generalized moonshine can be found in \cite{carnahan2018monstrous}. New relations between penumbral moonshine and generalized moonshine have also been found \cite{pmp, pmt}. These developments point towards the possibility of a unified theory of moonshine that would encompass all known examples. At the moment the connection of these results to physics is unclear, but as mentioned earlier, there are reasons to expect some kind of structure related to CFT or VOAs to lie behind these new connections.

\section{Open questions}
Moonshine mysteries, of mathematical and physical natures, abound. Chief among these mysteries is the missing umbral moonshine modules: while we have several isolated examples of umbral moonshine modules a uniform construction still eludes us, including a module for the $M_{24}$ moonshine that instigated these modern developments. Similarly, we do not yet have modules for the penumbral and O'Nan moonshines, and it is not yet known if and how these examples may be connected to physical ideas. More broadly, although all instances of moonshine enjoy some connection to a genus zero property, we do not yet understand the origin of these properties; one may hope that physics will ultimately provide a uniform answer to the question ``what is moonshine?'' Further, whether and how Mathieu and umbral moonshines are ultimately related to the geometry of K3 surfaces, or string theory on K3, is not yet understood, in spite of the context of the original observation of \cite{Eguchi:2010ej}. The symmetry surfing program has had some success, but further progress will likely require studying suitable connections on the bundle of K3 CFTs over their moduli space. It would also be interesting to develop a physical interpretation of the Rademacher summability properties of the McKay-Thompson series of umbral moonshine.

We may also ask a number of other questions raised by the aforementioned works: is there a generalization of the BKM algebras found in Monstrous and Conway moonshines in the case of the other moonshines? If so, is there a relationship to the BPS states of special string compactifications? Can we characterize the natural geometric or physical objects which have sporadic group symmetries? Are there instances of moonshine, or at least interesting sporadic group symmetries, that act on enumerative invariants of Calabi-Yau threefolds (or other spaces)? How could we understand such observations in a topological string? These questions, while already numerous, are far from exhaustive. Their eventual answers will doubtless shape our understanding of the mathematical foundations of string theory.

In spite of these mysteries, new connections between moonshine and other areas of physics are just beginning to be unearthed. We content ourselves with mentioning just a few new directions. One recent proposal \cite{Johnson-Freyd:2020itv} for understanding Mathieu moonshine takes place in the arena of algebraic topology, employing topological modular forms\footnote{Here, a natural possibility is that the $M_{24}$ symmetries may emerge when acting on a K3 surface with 24 points removed, as in $E_8 \times E_8$ small heterotic instantons. Preliminary explorations of Mathieu moonshine in the heterotic string on K3 include \cite{Harrison:2013bya}.}, and building on recent work which defines a new mock modular ``secondary elliptic genus'' as a natural invariant of 2d supersymmetric QFTs \cite{Gaiotto:2019gef}. Non-invertible symmetries and topological defect lines have been studied in the Monster VOA \cite{Lin:2019hks} and naturally capture its self-dualities. Moonshine groups have long been understood to be automorphism groups of classical error-correcting codes, but connections between non-chiral CFTs/sigma models to quantum error-correcting codes have been recently uncovered \cite{Harvey:2020jvu} and provide useful input to the modular bootstrap program \cite{Dymarsky:2020bps} \footnote{We also note that that modular representation theory has been recently employed to great effect in the modular bootstrap, providing another close connection between number theory and conformal field theory \cite{Kaidi:2021ent}.}. We anticipate that new connections between moonshine and other parts of string theory, field theory, and mathematics will continue to emerge.

\section*{Acknowledgements}
We thank S.~Carnahan, J.~Duncan, M.~Gaberdiel, G.~Moore, B.~Rayhaun, A.~Taormina and K.~Wendland for helpful comments on a draft version of this manuscript. The work of S.M.H. is supported by the National Science and Engineering Council of
Canada and the Canada Research
Chairs program. The work of J.H. is supported by National Science Foundation grant PHY-1520748. N.P. acknowledges support from the University of Washington and the DOE award DE-SC0022347.

\bibliographystyle{utphys}
\bibliography{references}

\end{document}